\documentclass{Interspeech2024}

\interspeechcameraready

\title{NTU-NPU System for Voice Privacy 2024 Challenge}
\name[affiliation={1,2}]{Nikita}{Kuzmin}
\name[affiliation={1}]{Hieu-Thi}{Luong}
\name[affiliation={3}]{Jixun}{Yao}
\name[affiliation={3}]{Lei}{Xie}
\name[affiliation={4}]{Kong Aik}{Lee}
\name[affiliation={1}]{\\Eng Siong}{Chng}

\address{
  $^1$Nanyang Technological University
  $^2$Institute for Infocomm Research, A$^\star$STAR \\
  $^3$Audio, Speech and Language Processing Group (ASLP@NPU) \\
  $^4$The Hong Kong Polytechnic University}
\email{S220028@e.ntu.edu.sg, hieuthi.luong@ntu.edu.sg, yaojx@mail.nwpu.edu.cn}

\usepackage{float}
\usepackage{multirow, tabularx}
\usepackage[title]{appendix}

\newcolumntype{P}[1]{>{\centering\arraybackslash}p{#1}}
\newcolumntype{M}[1]{>{\centering\arraybackslash}m{#1}}
\newcolumntype{L}{>{\raggedright\arraybackslash}X}


\begin{document}

\maketitle

\begin{abstract}
    In this work, we describe our submissions for the Voice Privacy Challenge 2024. Rather than proposing a novel speech anonymization system, we enhance the provided baselines to meet all required conditions and improve evaluated metrics. Specifically, we implement emotion embedding and experiment with WavLM and ECAPA2 speaker embedders for the B3 baseline. Additionally, we compare different speaker and prosody anonymization techniques. Furthermore, we introduce Mean Reversion F0 for B5, which helps to enhance privacy without a loss in utility. Finally, we explore disentanglement models, namely \ss-VAE and NaturalSpeech3 FACodec.
\end{abstract}

\section{Introduction}

Voice Privacy Challenge 2024 \cite{tomashenko2024voiceprivacy} introduces a new metric to evaluate the ability to maintain the conveyed emotion of the original source speaker.

Given the baselines provided by organizers, we test various techniques and methods aimed at improving the evaluated metrics.
Specifically, we create submissions for all four of EER conditions by using a modified version of B3 \cite{b3_system} and B5 \cite{champion2023anonymizing} systems, with the former responsible for conditions with EER lower than 30\% and the latter for those with EER from 30\% and above. Additionally, we experiment with disentanglement-based models such as \ss-VAE \cite{lu2023disentangled} and NaturalSpeech3 FACodec \cite{Ju2024NaturalSpeech3Z}.

The rest of the paper is constructed as follows: Section \ref{sec:baselines} summarizes the two baselines that we used for our submissions, Section \ref{sec:method} introduces all the modifications and proposed techniques used in our experiments, Section \ref{sec:experiments} provides the detailed results and the systems that we submitted for evaluation, Section \ref{sec:conclusion} concludes our findings. Furthermore, we provide the summary tables with a description of submitted systems in Appendix \ref{ap:summary_tables}.

\section{Baseline Systems}
\label{sec:baselines}

In this section, we summarize two baseline systems, B3 and B5, that we use in our experiments. A modified version of B3 is used to create submission for the first condition (EER$_1$) with EER between 10\% and 20\% and the second condition (EER$_2$) with EER between 20\% and 30\%, while a modified version of B5 is used for the third condition (EER$_3$) with EER between 30\% and 40\% and the last condition (EER$_4$) with EER larger than 40\%. The details about all proposed changes are laid out in Section \ref{sec:method}.
%In this section, we introduce the submitted systems, covering all four privacy requirements. We start by describing the systems used for the minimum EER$_1=10\%$ and EER$_2=20\%$ privacy scenarios. First group of these systems is based on Baseline 3 (B3), which employs speech synthesis on top of anonymized speaker embeddings and prosodic information while preserving linguistic content.

\subsection{B3}

The baseline system B3 uses a Wasserstein generative adversarial network with Quadratic Transport Cost (WGAN-QC) \cite{Liu2019WassersteinGW}  to generate artificial pseudo-speaker embeddings, anonymizing the speaker's identity through four main steps:
\begin{enumerate}
    \item \textbf{Phonetic Transcriptions Extraction:} Phonetic transcriptions are extracted using an end-to-end automatic speech recognition (ASR) model with a hybrid CTC-attention architecture.
    \item \textbf{Speaker Embeddings} Speaker embeddings are obtained using an adapted Global Style Tokens (GST) model \cite{wang2018style}. 
    \item \textbf{Anonymization:} The original speaker embedding is swapped with an artificial one generated by a WGAN. If the cosine distance between the artificial and original embeddings is less than 0.3, the replacement is considered successful. Otherwise, the process is repeated up to 30 times. Additionally, the pitch and energy values for each phoneme are adjusted using random values between 0.6 and 1.4.
    \item \textbf{Speech Synthesis:} The anonymized speaker embedding, adjusted prosody, and original phonetic transcription are used to create anonymized speech using the FastSpeech2 model and HiFi-GAN \cite{kong2020hifi} vocoder, as implemented in IMS-Toucan \cite{lux2023controllable}.
\end{enumerate}

\begin{figure*}[t]
\centering
\includegraphics[width=\linewidth]{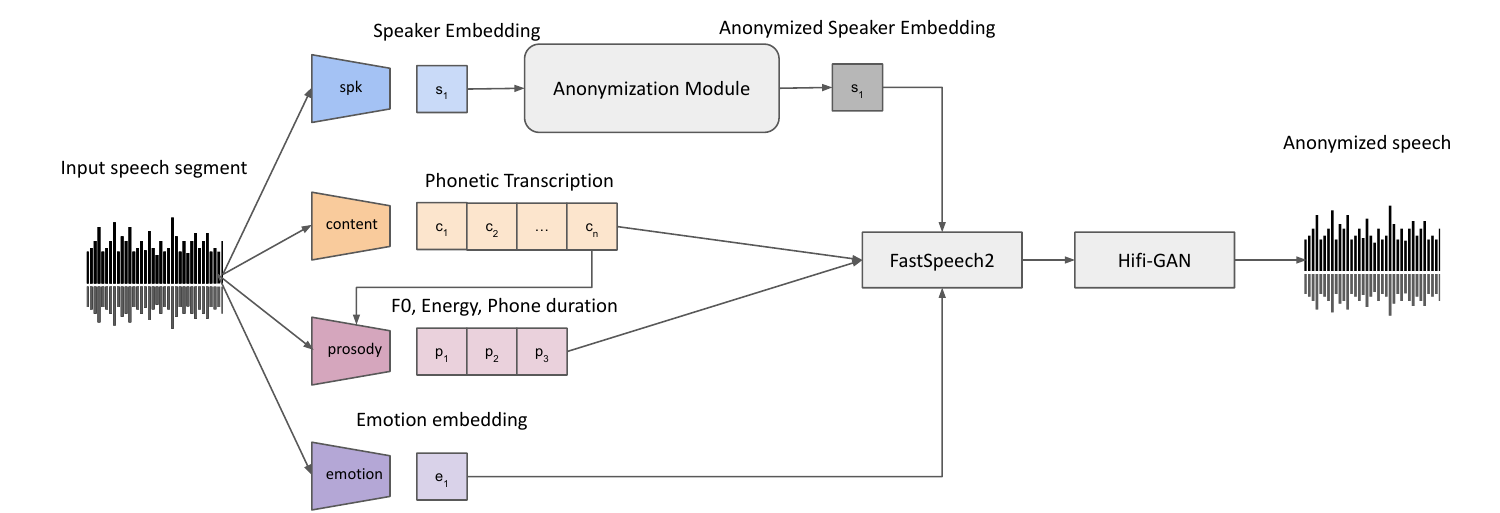}
\caption{Schematic diagram of Modified system B3.}
\label{fig:modified_b3_scheme}
\end{figure*}

\subsection{B5}

The B5 system used a HiFi-GAN model conditioned on fundamental frequency and a linguistic representation of the source utterance along with speaker embedding of a designated speaker to generate anonymized speech.
\begin{enumerate}
    \item \textbf{Fundamental Frequency (F0):} B5 uses a pytorch implementation of YAAPT Pitch Tracking \cite{zahorian2008spectral} to extract F0 from speech. In authors' of B5 thesis \cite{champion2023anonymizing}, the original authors of B5 suggest several complementary normalization or transformations to be applied to F0, none of which are included in B5. 

    \item \textbf{Linguistic Representation:} B5 uses the output of a vector quantization bottleneck layer (VQ-BN) put the top of the acoustic model (AM) of an automatic speech recognition (ASR) trained to classify left-biphone as the linguistic representation.

    \item \textbf{Speaker Embedding:} a designed speaker embedding of a speaker included in the training stage is used to change the voice of anonymized speech. We randomly pick a speaker embedding to anonymize each utterance similar to the B5 baseline provided by the organizer.
\end{enumerate}
We use the same pre-trained B5 model provided by organizers without doing any further training or tuning. Instead, we introduce a new method of transformation that can be applied to F0 in the inference stage, which is discussed in Section \ref{subsec:mean_reversion_f0}

\section{Our Methods}
\label{sec:method}
In this subsection, we elaborate on the details of proposed modifications to the baseline systems.

\subsection{Modifications of B3}

We experimented with the following main modifications for this system is as follows:
\begin{itemize}
    \item Emotion embeddings are implemented as an additional input to the FastSpeech2 model.
    \item The Global Style Tokens (GST) model is replaced by different speaker embedders such as WavLM \cite{Chen2021WavLM} and ECAPA2 \cite{ecapa2_paper}.
    \item Some experiments related to speaker selection strategy for anonymization and prosody manipulation.
\end{itemize}
    
    We start off with emotion embeddings. For extracting emotion-based embeddings, we employ a fine-tuned Wav2Vec2 Large Robust model \cite{10089511} on MSP-Podcast \cite{Lotfian2019BuildingNE}. Notably, the model is pruned from 24 to 12 transformers, and the CNN component is frozen prior to fine-tuning. 
    Embeddings are extracted from the hidden layer, which has 1024-dimensional vectors as output. We employ it to FastSpeech2 in the same way as speaker embeddings in \cite{meyer22_spsc} by adding one more linear projection and concatenating it with an output from Conformer \cite{conformergulati}.

    In addition to the GST model, we implement different speaker embedding models such as ECAPA2 and WavLM with 128 and 512 embedding sizes correspondingly. As both these speaker embedders work on audios instead of spectrograms, we add a pre-trained HiFi-GAN to the setup for the second phase of FastSpeech2 training in order to generate audio and extract embeddings for cycle consistency loss. In contrast to the GST model, ECAPA2 and WavLM speaker embedders are frozen during FastSpeech2 training. Similarly, the HiFi-GAN is also frozen.

    Furthermore, we explore various anonymization strategies, such as random speaker selection, which involves replacing the source speaker embedding for each utterance with a randomly selected embedding from a pool of embeddings. Additionally, we evaluated the importance of the usage of cross-gender for anonymization for the modified model. In cross-gender anonymization technique, we select a target speaker from the pool that has the opposite gender with respect to the source speaker. Finally, we examine how different powers of prosody anonymization affect privacy-based and utility-based metrics. To be more specific, we experiment with different offsets for pitch and energy multipliers. 

    The training process is the same as for the B3 baseline system. We retrain each part of the system except HiFi-GAN and ASR model.

\subsection{Disentanglement-based models}
Next, we explore disentanglement-based models, which might be useful for removing speaker-related information from other components (such as prosody, content, and acoustic information). We compare two models: ß-VAE [4] and NaturalSpeech3 FACodec [5]. Anonymization for these models is implemented by passing anonymized speaker embedding instead of source embedding and then performing voice conversion.

\subsection{Mean Reversion F0 for the B5 system}
\label{subsec:mean_reversion_f0}

\begin{figure*}[t!]
    \centering
    \includegraphics[width=0.85\linewidth]{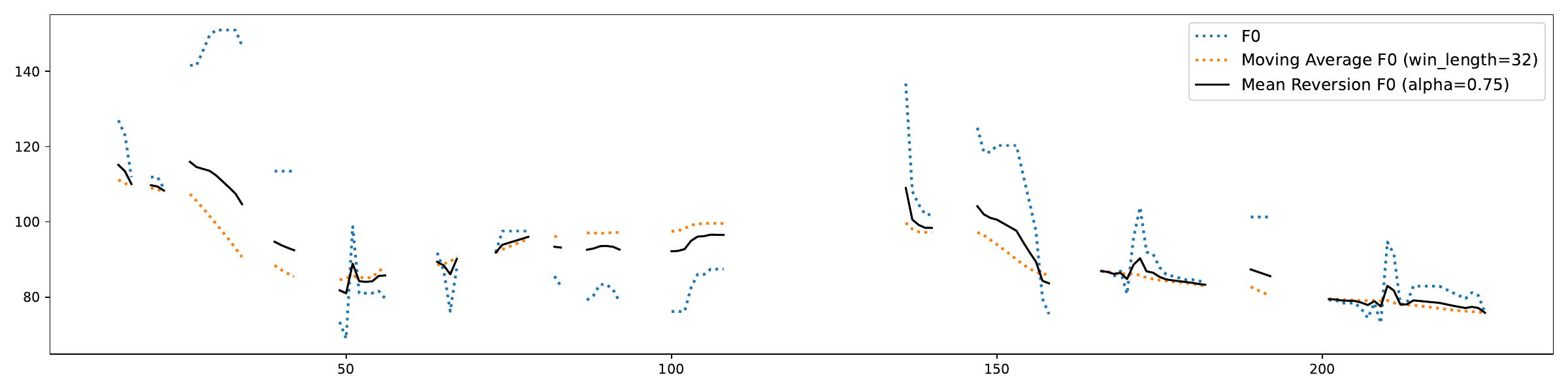} \\
    (a) Mean Reversion F0 \\ \vspace{3mm}

    \includegraphics[width=0.85\linewidth]{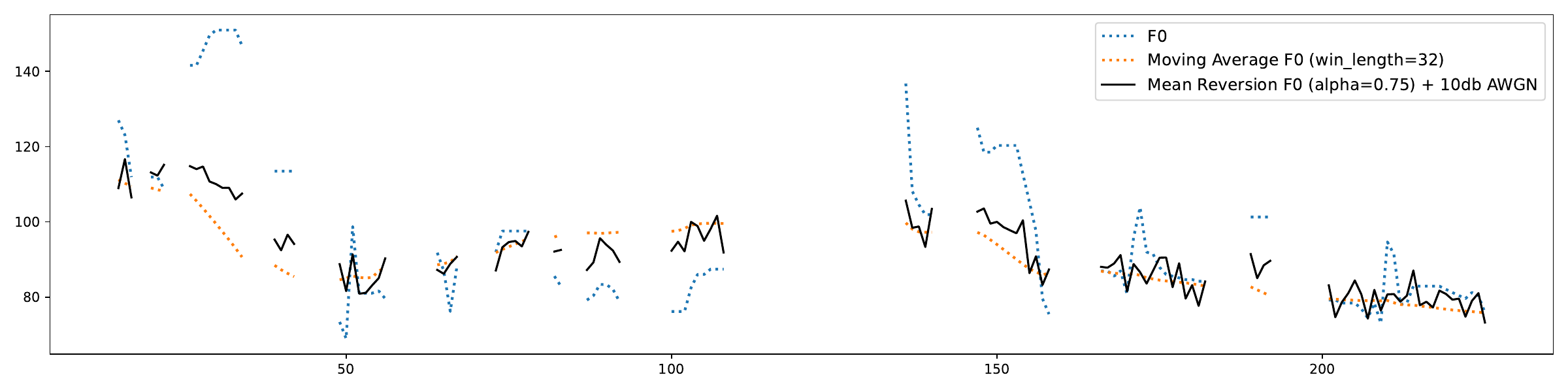} \\
    (b) Mean Reversion F0 with a 10-dB white gaussian noise
    \caption{Examples of Mean Reversion F0 with and without additive noise}
    \label{fig:mean_reversion_f0}
\end{figure*}

For conditions EER$_3$ and EER$_4$, we base our submissions on the B5 \cite{champion2023anonymizing} system provided by the organizer. Champion proposes a voice anonymization model using a HiFi-GAN vocoder which takes in the F0 and the linguistic representation of a source utterance and then turns it into the speech with the voice of a target speaker. In his thesis, the author explores different transformation techniques that can be applied to F0 including linear transformation, Additive White Gaussian Noise, and quantization. In this work, we propose a new type of transformation that uses the original and the n-frame moving average F0 ($\overline{F_0}$), with n=32 in our calculation:

\begin{equation}
    \hat{F_0} = (1-\alpha)F_0 + \alpha \overline{F_0} \\
\end{equation}
with $\alpha=0$, we get the original F0, and with $\alpha=1$, we get the moving average F0. For any $\alpha$ between 0 and 1, we obtain a mean reversion F0 which is a value weighted toward the mean. The motivation for this method is that we can reduce the dynamic range of F0, which is one characteristic of voice, and move it to a more neutral value. Moreover, the calculation is based on a short window instead of the entire utterance like the linear transformation method. Note that, we remove unvoiced frames when calculating the moving average F0.

We then apply an Additive White Gaussian Noise on top of mean reversion F0 to push up the EER. Figure \ref{fig:mean_reversion_f0} shows an example of the mean reversion F0 with and without additive noise.

\section{Experiments}
\label{sec:experiments}
In this section, we provide experimental results for the system modifications explained in Section \ref{sec:method}. To begin, we present the results for different modifications of system B3. The modified system B3 is illustrated in Figure \ref{fig:modified_b3_scheme}.

\subsection{Modified B3}
\begin{table*}[t!]
    \centering
    \footnotesize
    \caption{Comparison between systems with and without emotion embedding with different speaker embedder and prosody anonymization. $+$ in speaker anonymization column corresponds to Random-Speaker selection from LibriTTS-clean-100 for each source utterance. $+$ in prosody anonymization column corresponds to the systems with prosody multipliers from $[0.6, 1.4]$ range. }
    
    \begin{tabular}{c c c c | c c | c c | c c}
      \toprule
      Speaker & Speaker & Prosody & Emotion & \multicolumn{2}{c|}{EER} & \multicolumn{2}{c|}{UAR} & \multicolumn{2}{c}{WER} \\
       Anonymization & Embedder & Anonymization & Embedding & dev & test & dev & test & dev & test \\
      \hline
      --   & -- &  -- & --                            & 5.72   & 4.59   & 69.08 & 71.06 & 1.80 & 1.85 \\
      +         & GST & + & --                     & 25.76  & 28.42  & 37.97 & 37.39 & 4.33 & 4.33 \\
      \hline
      +   & GST  & + & +                       & 22.59  & 24.09  & 42.52 & 41.74 & 4.39          & 4.40 \\
      +   & GST  & -- & +                & 16.88  & 17.45  & 42.76 & 43.21 & \textbf{3.81} & \textbf{3.83} \\
      +  & WavLM & -- & +         & 17.97  & 16.64  & \textbf{43.84} & \textbf{45.67} & 4.54 & 4.69 \\
      +  & ECAPA2 & -- & +         & 19.48	& 22.55 & 42.53 & 42.37 & 4.83 & 4.80 \\
      \bottomrule
    \end{tabular}

  \label{tab:exp_emotion_emb}
\end{table*}
From the results in \ref{tab:exp_emotion_emb}, it can be observed that emotion embeddings help to improve Emotion Recognition performance while maintaining ASR performance at the same level. However, there is some degradation in privacy, which might be due to speaker identity leakage in the emotion embeddings. In addition, we provide experimental results for this system without prosody anonymization to check how modifications of prosody affect SER performance. As shown in the results, removing prosody modifications improves SER and ASR but also reduces privacy, making this system suitable for condition with a minimum EER$_1=10\%$.

% \begin{table}[H]
%     \setlength{\tabcolsep}{0.4em}
%     \footnotesize
%     \centering
%     \begin{tabular}{c | c c | c c | c c}
%       \toprule
%       Speaker & \multicolumn{2}{c|}{EER} & \multicolumn{2}{c|}{UAR} & \multicolumn{2}{c}{WER} \\
%        Embedder & dev & test & dev & test & dev & test \\
%       \hline
%       Original      & 5.72   & 4.59   & 69.08 & 71.06 & 1.80 & 1.85 \\
%       B3            & 25.24  & 27.32  & 38.09 & 37.57 & 4.29 & 4.35 \\
%       B3 (repro)    & 25.20  & 27.78  & 38.40 & 37.70 & 4.30 & 4.40 \\
%       \hline
%       \ss-VAE       & 25.02  & 27.83  & 39.70 & 35.30 & 6.47 & 6.37 \\
%       NS3           & 25.66  & 26.07  & 38.96 & 37.12 & 5.67 & 5.53 \\
%       ECAPA2        & 21.62  & 27.04  & 39.16 & 37.92 & 5.34 & 5.33 \\
%       WavLM         & 24.48  & 28.34  & 37.34 & 36.64 & 5.95 & 5.63 \\
%       \bottomrule
%     \end{tabular}
%   \caption{todo:rename this table}
%   \label{tab:exp_spk_embs}
% \end{table}

Table \ref{tab:exp_spk_selection} shows a comparison between WGAN and Random-Speaker anonymization techniques. There is almost no difference in the privacy and utility metrics for these methods, so we chose to stick with Random-Speaker as it requires no training.
\begin{table}[H]
    \centering
    \setlength{\tabcolsep}{0.55em}
    \footnotesize
      \caption{Comparison between WGAN anonymization strategy trained on LibriTTS-clean-100 and Random Speaker (Rnd-Spk) selection from LibriTTS-train-clean-100 per each source utterance.}
    \scalebox{0.95}{
    \begin{tabular}{c | c c | c c | c c}
      \toprule
      Anon &\multicolumn{2}{c|}{EER} & \multicolumn{2}{c|}{UAR} & \multicolumn{2}{c}{WER} \\
       Type & dev & test & dev & test & dev & test \\
      \hline
      WGAN                    & 25.20  & 27.78  & 38.40 & 37.70 & 4.30 & 4.40 \\
      \hline
      Rnd-Spk                 & \textbf{25.76}  & \textbf{28.42}  & 37.97 & 37.39 & 4.33 & 4.33 \\

      \bottomrule
    \end{tabular}}
  \label{tab:exp_spk_selection}
\end{table}

Next, Table \ref{tab:exp_prosody_multipliers} compares different ranges for multipliers involved in prosody anonymization. The results indicate that fewer prosody modifications result in worse privacy but better utility. This finding is useful for VPC2024 as it allows us to find a better trade-off between privacy and utility for specific EER conditions.

\begin{table}[H]
    \centering
    \setlength{\tabcolsep}{0.55em}
    \footnotesize
    \caption{Comparison between difference range for F0 and energy multipliers. The bottom row corresponds to the system without prosody manipulation.}
    \scalebox{0.92}{
    \begin{tabular}{c | c c | c c | c c}
      \toprule
       Multiplier &\multicolumn{2}{c|}{EER} & \multicolumn{2}{c|}{UAR} & \multicolumn{2}{c}{WER} \\
       Range & dev & test & dev & test & dev & test \\
      \hline
      $[0.6, 1.4]$                 & 25.76  & 28.42  & 37.97 & 37.39 & 4.33 & 4.33 \\
      \hline
      $[0.7, 1.3]$                 & 23.93  & 25.62  & 37.49 & 37.59 & 4.07 & 4.05 \\
      $[0.8, 1.2]$                 & 22.70  & 25.92  & 38.01 & 37.96 & 3.89 & 3.91 \\
      $[0.9, 1.1]$                 & 19.88  & 22.62  & 39.03 & 37.17 & 3.80 & 3.77 \\
      --                           & 19.47  & 21.82  & 38.91 & 38.11 & 3.70 & 3.75 \\

      \bottomrule
    \end{tabular}}
  \label{tab:exp_prosody_multipliers}
\end{table}

\subsection{Disentanglement-based models}
The comparison results between \ss-VAE and NaturalSpeech3 FACodec are shown in Table \ref{tab:exp_disentanglement_models}. It can be seen that \ss-VAE performs poorly in utility-based tasks, likely because of the fact that content representations are not rich enough. 

\begin{table}[t]
  \caption{Comparison between \ss-VAE and NS3 disentanglement models.}
  \label{tab:exp_disentanglement_models}
    \centering
    \footnotesize
    \setlength{\tabcolsep}{0.6em}
    \scalebox{0.95}{
    \begin{tabular}{c | c c | c c | c c}
      \toprule
      Model & \multicolumn{2}{c|}{EER} & \multicolumn{2}{c|}{UAR} & \multicolumn{2}{c}{WER} \\
       & dev & test & dev & test & dev & test \\
      \hline
      Original                      & 5.72   & 4.59   & 69.08 & 71.06 & 1.80 & 1.85 \\
      \midrule
      \ss-VAE                 & 10.71  & 10.49  & 30.38 & 31.28 & 67.72 & 65.5 \\
      NS3                     &  9.29  &  8.78  & 51.64 & 52.89 & 2.97  & 2.77 \\
      \bottomrule
    \end{tabular}}

\end{table}

As one might notice from the results in Table \ref{tab:exp_disentanglement_models}, NaturalSpeech3 has decent utility results. Therefore, we decided to employ anonymization techniques to improve privacy protection for NS3, aiming to meet a condition with minimum EER$_1=10\%$. We experimented with the following tricks: Additive White Gaussian Noise (AWGN) to Speaker Embedding and conversing a source speaker to a target speaker of the opposite gender (cross-gender). The results are shown in Table \ref{tab:exp_ns3_awgn_crossgender}. 
As we can see, cross-gender conversion helps to improve privacy and ASR performance on the corresponding test sets. Interestingly, it also improves SER performance on both development and test sets. As expected, AWGN enhances privacy at the cost of utility. 

Our results underscore the need for a balance between privacy and utility, as methods like AWGN and prosody anonymization can strengthen privacy but also impact system performance. This balance is essential for creating anonymization techniques that are both secure and functional.

\begin{table}[t]
    \centering
    \footnotesize
    \setlength{\tabcolsep}{0.4em}
    \caption{Comparison between NaturalSpeech3 FACodec systems with different power of AWGN applied to speaker embedding. The speaker anonymization module consists of averaging 100 embeddings randomly selected from a pool of 200 farthest embeddings (LibriTTS-train-clean-100) from source utterance by cosine scoring.}
    \scalebox{0.95}{
    \begin{tabular}{c c | c c | c c | c c}
      \toprule
      scale, & Cross & \multicolumn{2}{c|}{EER} & \multicolumn{2}{c|}{UAR} & \multicolumn{2}{c}{WER} \\
      $10e^{-3}$ & Gender & dev & test & dev & test & dev & test \\
      \hline
       -- &  --                            &  9.29  &  8.78  & 51.64 & 52.89 & 2.97  & 2.77 \\
      \hline
      75  & --                     & 12.25  & 9.14  & 48.00 & 48.09 & 4.66 & 4.63 \\
      75  & +                      & 12.09  & 10.46 & 49.20 & 49.12 & 4.97 & 4.60 \\
      78  & +                      & 12.42  & 10.24 & 49.10 & 49.39 & 5.44 & 5.07 \\
      80  & --                     & 12.63  & 9.42  & 47.33 & 48.35 & 5.37 & 5.40 \\
      80  & +                      & 13.66  & 10.10 & 48.82 & 48.95 & 5.69 & 5.37 \\
      90  & --                     & 12.41  & 10.45 & 47.61 & 47.10 & 7.04 & 6.45 \\
      \bottomrule
    \end{tabular}}
    \label{tab:exp_ns3_awgn_crossgender}
\end{table}

% \begin{table}[t]
%     \centering
%     \footnotesize
%     \setlength{\tabcolsep}{0.4em}
%     \caption{Comparison between NaturalSpeech3 FACodec systems with different power of AWGN applied to speaker embedding. The speaker anonymization module consists of averaging 100 embeddings randomly selected from a pool of 200 farthest embeddings (LibriTTS-train-clean-100) from source utterance by cosine scoring.}
%     \scalebox{0.95}{
%     \begin{tabular}{c c c | c c | c c | c c}
%       \toprule
%       Speaker & AWGN & Cross & \multicolumn{2}{c|}{EER} & \multicolumn{2}{c|}{UAR} & \multicolumn{2}{c}{WER} \\
%       Anon &  & Gender & dev & test & dev & test & dev & test \\
%       \hline
%        -- & -- &  --                            &  7.40  &  6.25  & 63.36 & 62.46 & 2.69  & 2.51 \\
%        \hline
%        + & -- &  --                            &  9.29  &  8.78  & 51.64 & 52.89 & 2.97  & 2.77 \\
%       + & +  & --                     & 12.25  & 9.14  & 48.00 & 48.09 & 4.66 & 4.63 \\
%       + & +  & +                      & 12.09  & 10.46 & 49.20 & 49.12 & 4.97 & 4.60 \\
%       \bottomrule
%     \end{tabular}}
%     \label{tab:exp_ns3_awgn_crossgender}
% \end{table}

\subsection{Modified B5}

\begin{table}[t]
  \caption{Evaluation results of B5 using Mean Reversion F0 with different values of $\alpha$ in inference stage}
  \label{tab:mean_reversion_f0}
    \centering
    \footnotesize
    \scalebox{0.95}{
    \begin{tabular}{c| cc | cc | cc}
      \toprule
      $\alpha$ & \multicolumn{2}{c|}{EER} & \multicolumn{2}{c|}{UAR} & \multicolumn{2}{c}{WER} \\
        & dev & test & dev & test & dev & test \\
      \hline
      0.00 & 31.64 & 31.36 & 39.18 & 38.24 & 4.79 & 4.44 \\
      0.25 & 32.13 & 32.03 & \textbf{39.61} & 38.38 & 4.74 & 4.54 \\
      0.50 & 33.48 & 34.08 & 38.60 & 37.34 & \textbf{4.62} & 4.54 \\
      \textbf{0.75} & \textbf{38.56} & 37.48 & 38.06 & 37.60 & 4.70 & 4.47 \\
      1.00 & 37.91 & \textbf{37.93} & 38.50 & \textbf{38.78} & 4.79 & \textbf{4.43}\\
      \bottomrule
    \end{tabular}}

\end{table}

\begin{table}[H]
  \caption{Evaluation results of B5 using Mean Reversion F0 ($\alpha=0.75$) and AWGN with different magnitude of noise in inference stage}
  \label{tab:mean_reversion_f0_awgn}
    \centering
    \footnotesize
    \scalebox{0.95}{
    \begin{tabular}{c| cc | cc | cc}
      \toprule
      dB & \multicolumn{2}{c|}{EER} & \multicolumn{2}{c|}{UAR} & \multicolumn{2}{c}{WER} \\
        & dev & test & dev & test & dev & test \\
      \hline
      0 & 38.56 & 37.48 & 38.06 & 37.60 & 4.70 & 4.47 \\
      5 & 39.58 & 40.00 & 38.91 & 37.12 & 4.67 & 4.49 \\
      \textbf{10} & 42.46 & \textbf{43.15} & \textbf{39.41} & \textbf{38.47} & \textbf{4.63} & \textbf{4.40} \\
      15 & \textbf{42.97} & 40.36 & 38.50 & 37.49 & 4.66 & 4.50 \\
      30 & 41.43 & 39.62 & 38.41 & 37.88 & 4.77 & 4.64 \\
      \bottomrule
    \end{tabular}}

\end{table}

Table \ref{tab:mean_reversion_f0} lists the results of the Mean Reversion F0 method discussed in Section \ref{subsec:mean_reversion_f0} given different $\alpha$ values. We can see a general trend that EER increases when $\alpha$ is increased while UAR and WER fluctuate but not very significantly. We submitted the sample generated with $\alpha=0.75$ for the condition EER$_3$.

For the last condition EER$_4$, we add a 10-db AWGN to the mean reversion F0 with $\alpha=0.75$ and manage to obtain an EER above 40\%. The result can be found in Table \ref{tab:submitted_systems}. 

We note that the EER results of these systems were highly volatile during our experiments, often producing different results even if we ran with the same configuration. It seems that convergence of an attacker ASV model depends on factors such as the machine, GPU, randomly picked speaker embedding, and other random parameters. The systems that we selected for submissions were based on the results available at that time and represented our methods of Mean Reversion F0 and AWGN.

\subsection{Submitted systems}
In this subsection, we provide a summary of all submitted systems. Table \ref{tab:submitted_systems} shows privacy and utility results for each of the conditions.
\begin{table}[H]
    \centering
    \footnotesize
    \setlength{\tabcolsep}{0.4em}
    \caption{Results summary for all submitted systems grouped by achieved privacy conditions.}
    \scalebox{0.92}{
    \begin{tabular}{c c| c c | c c | c c}
      \toprule
      Condition & System & \multicolumn{2}{c|}{EER} & \multicolumn{2}{c|}{UAR} & \multicolumn{2}{c}{WER} \\
       & ID & dev & test & dev & test & dev & test \\
      \hline
      EER$_1$     &  1a                & 12.09   & 10.46   & 49.20 & 49.12 & 4.97 & 4.60 \\
                  &  1b                & 16.88   & 17.45   & 42.76 & 43.21 & 3.81 & 3.83 \\
      \hline
      EER$_2$     &  2a                & 21.47   & 24.13   & 44.67 & 42.78 & 4.21 & 4.29 \\
                  &  2b                & 20.07   & 22.85   & 39.18 & 37.67 & 3.61 & 3.68 \\
      \hline
      EER$_3$     &  3                & 38.56   & 37.48   & 38.06 & 37.60 & 4.70 & 4.47 \\
      \hline
      EER$_4$     &  4                & 42.46   & 43.15   & 39.41 & 38.47 & 4.63 & 4.40 \\
      \bottomrule
    \end{tabular}}
  \label{tab:submitted_systems}
\end{table}

Additionally, we prepared the tables in the Appendix \ref{ap:summary_tables} with a summary of architecture, input, output values, and training data for components in submitted systems.

\section{Conclusion}
\label{sec:conclusion}
In this system description, we modified the baseline systems (B3 and B5) for the Voice Privacy Challenge 2024 to enhance speaker anonymization while preserving emotional and content features. Specifically, we integrated emotion embeddings and different speaker embedders such as WavLM and ECAPA2 into system B3. In addition, we explored random-speaker and cross-gender anonymizations and different setups of prosody manipulation. For B5, we introduced Mean Reversion method and AWGN for prosody which allowed us to enhance privacy while maintaining utility.
Finally, we experimented with disentanglement-based approaches such as \ss-VAE and NaturalSpeech3. An additional analysis on NaturalSpeech3 showed promising results for the Voice Privacy problem.

% \begin{table}[th]
%   \caption{This is an example of a table}
%   \label{tab:example}
%   \centering
%   \begin{tabular}{ r@{}l  r }
%     \toprule
%     \multicolumn{2}{c}{\textbf{Ratio}} & 
%                                          \multicolumn{1}{c}{\textbf{Decibels}} \\
%     \midrule
%     $1$                       & $/10$ & $-20$~~~             \\
%     $1$                       & $/1$  & $0$~~~               \\
%     $2$                       & $/1$  & $\approx 6$~~~       \\
%     $3.16$                    & $/1$  & $10$~~~              \\
%     $10$                      & $/1$  & $20$~~~              \\
%     \bottomrule
%   \end{tabular}
  
% \end{table}

% \subsection{Equations}

% Equations should be placed on separate lines and numbered. We define
% % 
% \begin{align}
%   x(t) &= s(t') \nonumber \\ 
%        &= s(f_\omega(t))
% \end{align}
% % 
% where \(f_\omega(t)\) is a special warping function. Equation \ref{equation:eq2} is a little more complicated.
% % 
% \begin{align}
%   f_\omega(t) &= \frac{1}{2 \pi j} \oint_C 
%   \frac{\nu^{-1k} \mathrm{d} \nu}
%   {(1-\beta\nu^{-1})(\nu^{-1}-\beta)}
%   \label{equation:eq2}
% \end{align}
% 

\section{Acknowledgements}
The computational work for this article was partially performed on resources of the National Supercomputing Centre, Singapore.

\bibliographystyle{IEEEtran}
\bibliography{mybib}

\appendix

\section{Summaries for submitted systems}
\label{ap:summary_tables}
\begin{table*}[h!]
    \caption{Description of Systems 1a.}

    \begin{tabular}{|c|M{2cm}|p{8cm}|p{2cm}|p{2cm}|}
      \hline
      \# & Module & Description & Output features & Data \\
      \hline
      1 & Encoder \cite{kumar2024high} & 4 Downsampling Convolution-based Layers \newline with Snake activation function \newline Input: speech waveform & Output vector$^{256}$ & Librilight train \cite{2019arXiv191207875K} \\
      \hline
      2 & Prosody extractor & Factorized Vector Quantization with 1 quantizer, codebook size: 1024 & Prosody vector$^{256}$ & Librilight train \\
      \hline
      3 & Content extractor & Factorized Vector Quantization with 2 quantizers, codebook size: 1024 & Content vector$^{256}$ & Librilight train \\
      \hline
      4 & Speaker embedding extractor & Several Conformer blocks &  Speaker embedding$^{256}$ & Librilight train \\
      \hline
      5 & Speaker anonymization module & Averaged 100 embeddings randomly selected from a pool of 200 farthest embeddings from source by cosine scoring \newline AWGN with scale$=0.075$ \newline Cross-gender & Anonymized speaker embedding$^{256}$ & LibriTTS: \newline train-clean-100 \\
      \hline
      6 & Decoder \cite{kumar2024high}                         & Upsampling Convolution-based Layers \newline with Snake activation function & speech \newline waveform & Librilight train \\
      \hline
    \end{tabular}
    \label{sum:sys_1a}
\end{table*}

\begin{table*}
    \caption{Description of Systems 1b, 2a, 2b.}
    \begin{tabular}{|c|M{2cm}|p{8cm}|p{2cm}|p{2cm}|}
      \hline
      \# & Module & Description & Output features & Data \\
      \hline
      1 & Prosody extractor & Phone aligner: 6-layer CNN + LSTM with CTC loss \newline F0 estimation using Praat \newline F0, energy, durations normalized by each vector’s mean & F0$^1$, energy$^1$ \newline phone durations$^1$ & LibriTTS: \newline train-clean-100 \\
      \hline
      2 & ASR & End-to-end with hybrid CTC-attention \newline Input: log mel Fbank$^{80}$ \newline Encoder: Branchformer \newline Decoder: Transformer \newline CTC and attention criteria & phonetic transcript with pauses and punctuation & LibriTTS: \newline train-clean-100 \newline train-other-500 \\
      \hline
      3 & Speaker embedding extractor & GST, trained jointly with SS model \newline Input: mel spectrogram$^{80}$ \newline 6 hidden layers + 4-head attention & GST speaker embedding$^{128}$ & LibriTTS: \newline train-clean-100 \\
      \hline
       4 & Emotion embedding extractor & \textbf{1b, 2a:} Dimensional Speech Emotion Recognition Model based on Wav2vec 2.0 \newline Input: Wav2vec 2.0 Large features & emotion embedding$^{1024}$ & MSP-Podcast (v1.7)\\ \cline{3-5}
         &     & \textbf{2b:} --                                                                                 & -- & -- \\
      \hline
      5 & Prosody modification module & \textbf{1b, 2b:} -- & -- & -- \\ \cline{3-5}
        &                             & \textbf{2a:} Value-wise multiplication of F0 and energy with random values in [0.7, 1.3)  & F0$^1$, energy$^1$ & LibriTTS: \newline train-clean-100 \\
      \hline
      6 & Speaker anonymization module & \textbf{1b:} Averaged 100 embeddings randomly selected from a pool of 200 farthest embeddings from source by cosine scoring + cross-gender  \newline \textbf{2a, 2b:} Random Speaker selection per each source utterance + cross-gender  & Anonymized speaker embedding$^{128}$ & LibriTTS: \newline train-clean-100 \\
      \hline
      7 & SS model & IMS Toucan implementation of FastSpeech2 \newline Input: F0$^1$ + energy$^1$ + phone durations$^1$ + phonetic transcript + GST embeddings$^{128}$ (\textbf{1b, 2a:} + emotion embeddings$^{1024}$) \newline Training criterion defined in FastSpeech2 & mel spectrogram$^{80}$ & LibriTTS: \newline train-clean-100 \\
      \hline
      8 & Vocoder & HiFi-GAN vocoder \newline Input: mel spectrogram$^{80}$ \newline Training criterion defined in HiFi-GAN & speech waveform & LibriTTS: \newline train-clean-100 \\
      \hline
    \end{tabular}
    \label{sum:sys_1b_2ab}
\end{table*}

\begin{table*}
    \caption{Description of Systems 3 and 4.}
    \begin{tabular}{|c|M{2cm}|p{8cm}|p{2cm}|p{2cm}|}
      \hline
      \# & Module & Description & Output features & Data \\
      \hline
      1 & F0 extractor & F0 extracted with s pytorch implementation of YAAPT \newline \textbf{3:} Using Mean Reversion F0 ($\alpha=0.75$) in inference \newline \textbf{4:} Using Mean Reversion F0 ($\alpha=0.75$) and 10-db AWGN & F0 & N/A \\ \hline 
      2 & ASR AM with VQ & Acoustic Model trained to identify left bi-phones \newline and a VQ bottleneck layer & Linguistic \newline representation & VoxPopuli \newline Librispeech: train-clean-100\\ \hline
      3 & Speaker embedding & One-hot vector represented speaker in training set & Speaker \newline embedding & LibriTTS: train-clean-100 \\ \hline
      4 & Speech Synthesis & HiFi-GAN vocoder \newline Input: F0 + linguistic representation + speaker embedding & Speech \newline waveform & LibriTTS: train-clean-100\\
      \hline
    \end{tabular}
    \label{sum:sys_34}
\end{table*}

\end{document}